\documentstyle[prl,aps]{revtex}
\begin{document}
\draft

\title{A Method for Detecting Possible Non-determinism in
a Time Series}

\author{D. D. Dixon}
\address{
Institute for Geophysics and Planetary Physics\\
University of California \\
Riverside, CA 92521\\
DIXON@UCRPH0.UCR.EDU
}
\author{M. Zak}
\address{
Jet Propulsion Laboratory\\
California Institute of Technology\\
Pasadena, CA, 91109
}
\author{J. P. Zbilut}
\address{
Department of Physiology\\
Rush Medical College\\
1653 W. Congress Pkwy.\\
Chicago, IL. 60612
}

\date{\today}

\maketitle

\begin{abstract}
A method for detecting possible non-deterministic dynamics
underlying a time series is introduced.  Non-deterministic dynamics
may arise
due to the failure of the Lipschitz condition in the equations
of motion.  At a singular point,
the phase
space trajectory of the system may jump from one solution to another.
This discontinous change implies divergence of the second derivative
of the solution whenever it passes near the singular point.  A time
series can be examined for such divergences, which may indicate
non-determinism in the dynamical system.  Examples with both simulated
and actual data are given.
\end{abstract}


\paragraph{Non-deterministic dynamics}

The standard approach to classical dynamics assumes a Laplacian point
of view, i.e., that the time evolution of a system is uniquely
determined by it's initial conditions~\cite{arnold}.  This view of
determinism results from the {\em Existence and Uniqueness Theorem}
of differential equations~\cite{coddington}, which requires that
the equations of motion everywhere satisfy the Lipschitz condition.
It has long been tacitly assumed that nature (in the classical
realm) is deterministic, and that correspondingly, the equations of
motion describing physical systems are Lipschitz.\\

However, there is no {\it a priori} reason to believe that nature
is Lipschitzian.  For instance, it has been shown\cite{zak1} that in
the case of a whip snapping, the physical solutions correspond
to equations of motion that violate the Lipschitz condition.  A
similar effect is seen when seismic waves approach the surface
of the earth~\cite{zak2}.  In
this paper, we are concerned with a particular implication of
non-Lipschitz equations of motion, namely, the possibility of
non-unique solutions.  If a dynamical system is non-Lipschitz at
a singular point, it is possible that several solutions will intersect at
this point~\cite{zak3,dixon1}.  Since this singularity is a common
point among many trajectories, the dynamics of the system {\em after}
the singular point is intersected is not in any way determined
by the dynamics before, hence the term {\em non-deterministic
dynamics.}  We hasten to emphasize, that the term
``non-deterministic'' should
not necessarily be construed to mean purely random.
On the contrary, careful
analysis demonstrates unique, and complicated
dynamics~\cite{zak3,dixon1,dixon2}, some of
which we describe here.\\

Of consequence to us in this paper is in possibility of
{\em non-deterministic chaos}.
For a non-deterministic system, it is entirely possible
(if not likely), that as the various solutions move away from
the singularity they will evolve very differently, and tend to diverge.
Several solutions coincide at the non-Lipschitz singularity,
and therefore
whenever a phase space trajectory comes near this point any arbitrarily
small perturbation may push the trajectory on to a completely different
solution.  As ``noise'' is intrinsic to any physical system, we
expect the time evolution of
a non-deterministic dynamical system to consist of a series of
transient trajectories, with a new one being chosen randomly
whenever the solution (in the presence of noise) nears the
non-Lipschitz point.  We term such behavior non-deterministic
chaos.\\

\paragraph{Example of non-deterministic chaos}

A physically motivated non-deterministic set of
equations comes from a simple model describing the dynamics
of neutron star magnetic fields~\cite{cummings}.  The model
envisions a neutron star to contain two spherical components,
each carrying opposite charge.  These charged spheres are allowed
to rotate differentially, and interact both mechanically and
electromagnetically.  The mechanical interaction is a simple
damping, taken to be proportional to the difference in the
two angular velocities.  The electromagnetic interactions
are a standard gyromagnetic term, which induces precession
of the magnetic moments, and the Landau-Lifshitz magnetic
damping~\cite{landau}, which tends to align the magnetic moments.\\

When appropriately scaled and expressed in cyclindrical
polar coordinates, the equations of motion of the magnetic
moment due to the rotation of one sphere take the form
\begin{eqnarray}\label{eq:nstar}
\dot{\rho} & = & \frac{\rho z}{\rho^2 + z^2} - \epsilon \rho,\nonumber\\
\dot{z} & = & \frac{z^2}{\rho^2 + z^2} - \epsilon z + \epsilon - 1,
\end{eqnarray}
where $\rho$ and $z$ are the scaled radial and $z$ components of
the magnetic moment, and $\epsilon$ is a parameter describing
the relative strengths of the mechanical and Landau-Lifshitz damping.
The solutions of eqs.\ \ref{eq:nstar} are a series
of closed loops, all sharing a common tangent point at the
origin~\cite{dixon2}.  In the presence of external
fluctuations, eqs.\ \ref{eq:nstar} should exhibit
non-deterministic chaos.  A numerically simulated time series
for eqs.\ \ref{eq:nstar} is shown in Fig.\ \ref{fig:nstar}.\\

For the above example, the non-Lipschitz singularity occurs
at a non-equilibrium point.  Other examples where the non-Lipschitz
behavior is seen at an equilibrium point have been
previously given~\cite{zak3}.  In such cases, time variation of
the stability of the equilibrium point may lead to
non-deterministic chaos~\cite{zak3,chen}.\\

\paragraph{Detecting non-determinism in time series}

Suppose for a moment that the time series of Fig.\ \ref{fig:nstar}
is a data set, for which we know nothing of the underlying
dynamical system.  Several approaches have been proposed
to detect determinism in an otherwise random appearing
data set.  However, such approaches {\em assume} that nature
is continuously differentiable, which is clearly not necessarily
the case.  It has been shown~\cite{crutchfield}
that such assumptions may lead one to believe that the
system is more complex than in actually is.\\

It is thus important to be able to detect non-determinism
in data, so that the {\em apparent} complexity is not
mistaken for actual complexity in the underlying physical
system.  The key feature of non-deterministic chaos is
the abrupt change of the trajectory as it jumps from one
solution to another every time it nears the singularity.
Treating this jump as essentially instantaneous, we might
write the solution
\begin{equation}
x(t) = \Theta(t - t_0) x_1(t) + \Theta(t_0 - t) x_2(t),
\end{equation}
where $\Theta(x)$ is the unit step function, and $t_0$ is the
time when the jump occurs.  If $x_1(t)$ and $x_2(t)$ are
non-trivial and different solutions, then some time derivative
of $x(t)$ will diverge at $t_0$.  Indeed, analysis of the
non-Lipschitz nature of the equations of motion indicate that
the divergence would be seen in the second (or higher) time
derivative.\\

\paragraph{Numerical example}

We take our numerical example from the data set shown in
Fig.\ \ref{fig:nstar}.  This time series was generated by
4th order Runge-Kutta integration of eqs.\ \ref{eq:nstar} with
a step size of 0.0001.  At each integration step, a Gaussian
random number with zero mean and standard deviation of
$10^{-8}$ was added to the dependent variables.  To mimic
the effect of finite sampling, only every 100th data point was
kept.  The second time derivative of $z(t)$ was computed
numerically using Lagrangian interpolation, and is shown in
Fig.\ \ref{fig:d2nstar}.  The divergences are plainly
visible, corresponding to positive slope zero crossings of the signal.
Such divergences are indicative of a non-deterministic jump in
the trajectory, and indeed, study of eqs.\ \ref{eq:nstar}
indicate that the singular behavior is expected at this
point~\cite{dixon2}.\\

\paragraph{Application to physiologic data}

It has been previously suggested~\cite{zbilut} that non-determinism
may appear in biological systems.  We believe that such dynamics
may be especially important in the
case of biologic data which have a close affinity with physical
models such as oscillations. Although such models are especially
informative, questions of control are not easily resolved. We
suspect non-deterministic chaos may prove to be an important
additional feature in such models.  In particular, as an infinite
number of solutions are accesible by an arbitrarily small perturbation
in the neighborhood of the singularity, selection of a particular solution
by a control mechanism is very easy.  The result would be a
series of nearly identical oscillations, where slight differences in
successive oscillations are randomly noise induced.  Again, the
derivative test may be able to detect this ``controlled''
non-deterministic chaos.\\

We have obtained the digitized records( 500 Hz) of the
oscillatory motion of a human forearm swinging at the elbow
extended out from the shoulder.  Instructions were given to the
subject to maintain the oscillations at a frequency maintained
by a metronome at approximately 1 Hz.
Note that high frequency noise will cause large fluctuations in the
time derivatives, and is unavoidable
due to the physiologic nature of the system. Because
the usual requirements of stationarity for signal averaging
could not be assumed, the data were smoothed by boxcar
averaging with a 5 point wide window applied 100 successive times.
We note that this procedure is not optimal in that any filtering procedure
will ``smooth out'' any divergence of higher order derivatives.  This is
equivalent to the well-known problems associated with the impulse
response functions of digital filters. Clearly, sharp divergences will show
up as wave functions of varying magnitude and phase dependent upon
local conditions.  Indeed, we suspect one reason why such divergences
have not been routinely encountered has been due to the ubiquitous need
to filter experimental data.
Once the signal has been smoothed, Lagrangian interpolation is used to
calculate the second derivative, shown in Fig.\ \ref{fig:arm}.
The results, while far from conclusive, are suggestive.
The ``divergences'', while not the largest oscillations, do bisect
the original oscillations quite nicely, which might
be expected given previous modeling results~\cite{eisenhammer,hubler}.
Further, the ``divergences'' show the correct sign, and while other
large oscillations in the second derivative are generally associated
with visible departures from smoothness in the original signal,
the signal around the extrema is fairly well behaved, and thus
the features in the second derivative associated with the extrema
of the original signal may indicate non-deterministic behavior
at these points.\\

Again, given the complications involved in examining this data
set, we caution against any conclusive interpretation.  However
the results seem compelling enough to warrant further investigation.\\

\paragraph{Conclusions}

Non-determinism may occur in a dynamical system that does not
satisfy the Lipshitz condition at a point or set of points.
In the presence of arbitrarily small fluctuations, this
non-determinism may give rise to behavior that appears very
complex.  We have suggested that, given the expected piecewise
continuous nature of non-deterministic chaotic signals, one
possible method of detecting this phenomenon in time series
is to look for divergences in some time derivative.  Application
to numerically simulated data has met with marked success.  The
derivative test, however, is limited by noise and other factors,
and will be most successful on ``clean'' data.  Appropriate
smoothing and/or signal averaging may mitigate these problems.\\

\paragraph{Acknowledgements}

We thank M. Latash for the arm data.
D.D.D. wishes to thank J. Wudka for many helpful
discussions.  M.Z. acknowledges support from the Agencies of the
U.S. Department of Defense through an agreement with the National
Aeronautics and Space Administration.

\begin{figure}
\caption{An example of non-deterministic chaos, showing
$y$ vs. $t$ for the numerically integrated
neutron star equations in the presence of noise.}
\label{fig:nstar}
\end{figure}

\begin{figure}
\caption{A plot of the time series of Fig.\
\protect\ref{fig:nstar}, along with
it's numerically calculated second time derivative.  The plot
of the second derivative has been scaled and shifted upward for clarity.
Note the divergences in the second derivative which occur
at positive slope zero crossings, corresponding to the location
of the singular point in the neutron star model.}
\label{fig:d2nstar}
\end{figure}

\begin{figure}
\caption{A portion of the unsmoothed arm data (dashed curve) and
the numerical second derivative of the smoothed data (solid curve).
The vertical dotted lines correspond to the expected location
of singular points in a non-deterministic model of human arm motion,
showing a clear correspondence with peaks in the second derivative
which are consistent with the expected behavior of a smoothed
delta function.}
\label{fig:arm}
\end{figure}

\end{document}